\documentclass[reprint,aps,twoside,prd,showkeys,showpacs,superscriptaddress]{revtex4-2}
\usepackage{epsfig}
\usepackage{amsmath}
\usepackage{amssymb}
\usepackage{txfonts}
\usepackage{bm}
\def\<{\left\langle}
\def\>{\right\rangle}
\begin{document}%\draft{}
\pagestyle{myheadings}
\markboth{{\rm PRST-AB~\underline{3}}\hfill
{\rm J\"URGEN STRUCKMEIER}\hfill 034202 (2000)}%
{{\rm PRST-AB~\underline{3}}\hfill{\rm STOCHASTIC EFFECTS IN REAL AND
SIMULATED \ldots\hfill 034202 (2000)}}
\title{Stochastic effects in real and simulated charged particle beams}
\author{J\"urgen Struckmeier}
\affiliation{Gesellschaft f\"ur Schwerionenforschung (GSI),
Planckstr.~1, 64291~Darmstadt, Germany}
%\date{\today}
\date{Received 6 December 1999; published 24 March 2000}
\begin{abstract}
The Vlasov equation embodies the smooth field approximation
of the self-consistent equation of motion for charged particle beams.
This framework is fundamentally altered if we include the
fluctuating forces that originate from the actual charge granularity.
We thereby perform the transition from a reversible description
to a statistical mechanics' description covering also the
irreversible aspects of beam dynamics.
Taking into account contributions from fluctuating forces
is mandatory if we want to describe effects like intrabeam
scattering or temperature balancing within beams.
Furthermore, the appearance of ``discreteness errors'' in computer
simulations of beams can be modeled as ``exact'' beam dynamics
that is being modified by fluctuating ``error forces''.
It will be shown that the related emittance increase
depends on two distinct quantities: the magnitude of the fluctuating
forces embodied in a friction coefficient $\gamma$, and the
correlation time dependent average temperature anisotropy.
These analytical results are verified by various computer simulations.
\end{abstract}
\maketitle
%\begin{multicols}{2}
\section{Introduction}\label{sec:intro}
Analytical approaches to the dynamics of charged particle beams
that are based on the Liouville --- or equivalently on the Vlasov ---
equation do not include effects due to the actual charge granularity.
A variety of beam phenomena are adequately described by this
continuous description.
As the first example, we cite the pioneering work of
I.M.~Kapchinskij and V.V.~Vladimirskij~\cite{kapvla} covering
the description of beam transport under space charge conditions.
As a second example, we may quote the well-understood transient effects
that occur if a beam is launched with a non-self-consistent phase space
density profile~\cite{strklarei,wangler,reiser}.
Furthermore, the various kinds of parametric resonances and
instabilities that may occur in the course of beam propagation
through focusing lattices and storage rings have been
successfully tackled on the basis of a perturbation analysis
of the Vlasov equation~\cite{hola,hotemp,birdsall,neil,strrei}.

Despite all these achievements, there is still an important
class of beam phenomena the analysis of which leads beyond
the scope of the Vlasov approach.
Due to the invariance of Vlasov's equation with
respect to time reversal~\cite{hobson}, we must realize that it
restricts the analysis to only reversible aspects of beam dynamics.
However, a reversible, continuous description of beam dynamics
no longer applies if the individual interactions of the point charges
must be taken into account.
Effects of elastic Coulomb scattering like the well-known phenomenon
of intrabeam scattering~\cite{piwi} observed for intense beams that
circulate in storage rings, or the process of temperature balancing
within a charged particle beam --- commonly referred to as
beam equipartitioning --- fall into this category.
In order to include these irreversible effects into our analytical
description of beams, the Vlasov approach must be generalized
appropriately~\cite{chandra,biso,struck-pa,struck}.
This will be achieved by switching from a deterministic to a
statistical treatment of beam dynamics, namely by separating the
actual forces that act on the beam particles into a smooth and a
fluctuating component.
We will review this transition in detail in Sec.~\ref{sec:langevin}.

Owing to the fact that an analytical solution for the problem of
particles interacting by Coulomb forces does not exists,
computer simulations have become the tool of choice for the
study of charged particle beams.
In these studies, the actual beam is represented by an ensemble of
simulation particles.
A simulation thus means to numerically integrate the coupled set of equations
of motion constituted by the particle ensemble and the beam optical lattice.
Although the equations of motion of individual particles are
invariant with respect to time reversal, the evolution of the
particle ensemble is inevitably rendered irreversible because
of the limited accuracy of numerical methods.
Therefore, a simulation based on individual particles can never be
a strict realization of a solution of the associated Vlasov equation.

The idea pursued in this article is to describe these ``computer noise''
effects analogous to random force effects emerging within
the granular charge distribution of a ``real'' beam.
We can then interpret the beam simulation results
within the framework of the generalized Vlasov approach, which will
be reviewed in Secs.~\ref{sec:fokker-planck} to~\ref{sec:fpegr}.
This allows us to separate effects caused by the specific
realization of the computer simulation from the ``real beam'' physics.
The onset of irreversibility in a computer simulation of a charged
particle ensemble will be demonstrated in Sec.~\ref{sec:iniemi}.
We numerically transform a beam forward a specific amount of time,
followed by the backward transformation to its starting point.
The reversible transient effect of ``initial emittance growth''
--- occurring for beams with non-self-consistent phase space densities,
as described by the Vlasov equation --- is rendered
irreversible because of the accumulated action of ``error'' forces
that inevitably accompany our numerical calculations.

In a second simulation example presented in Sec.~\ref{sec:long-term},
the joint occurrence of reversible and irreversible effects
within simulated charged particle beams is visualized.
For a specific time span after a numerical time reversal,
the beam evolution behaves reversible.
After this, the irreversible ``computer noise'' effects prevail,
indicated by a sharp change of the sign of the emittance growth rate.

In Sec.~\ref{sec:simutab}, we will
analyze the numerical emittance growth factors obtained
for different focusing lattices, matching conditions, and number
of simulation particles.
It will be shown that the specific emittance growth
rates emerging in these simulations can indeed be explained
within the framework of the generalized Vlasov approach.
We also investigate the scaling of these emittance growth factors
with the number of particles used in the simulation in order to
distinguish ``computer noise'' related effects from those occurring
within a real beam.
\section{Langevin Equation}\label{sec:langevin}
We start our analysis reviewing the single particle equation
of motion for a set of charged particles interacting through
Coulomb forces within the co-moving beam frame
\begin{equation}\label{deteq}
m\frac{d^{2}}{dt^{2}} \boldsymbol{x} -
\boldsymbol{F}_{{\rm ext}}(\boldsymbol{x},t) -
q\boldsymbol{E}_{{\rm sc}}(\boldsymbol{x}) = 0\,,
\end{equation}
with $m$ the particle mass and $q$ its charge,
$\boldsymbol{F}_{{\rm ext}}$ denoting the external force field and
$\boldsymbol{E}_{{\rm sc}}$ the total electric self-field generated by
all other particles.
If a total of $N$ particles of the same species is given, the $N$-body
distribution function
\begin{equation}\label{distri}
\rho = \rho(\boldsymbol{x}_{1},\boldsymbol{v}_{1}, \ldots,
\boldsymbol{x}_{N},\boldsymbol{v}_{N},t_{0})
\end{equation}
contains the complete information on the state of the system.
Eq.~(\ref{deteq}) together with the knowledge of (\ref{distri}) defines a
``reversible'' system, hence a system that is completely determined and
does not contain any sources for loss of information.
Without such losses, the system can be transformed to any instant
of time back and forth.
In principle, all effects occurring in charged particle beams can be
extracted from the time integration of Eq.~(\ref{deteq}).

Nevertheless, this picture is not adequate for the description
of real $N$-body systems if $N$ is very large, since the condition
that the initial $\rho$ is precisely known can never be fulfilled.
In addition, the detailed knowledge of $\rho(t)$ is usually not useful.
Therefore, a statistical description of the time evolution of the particle
ensemble is appropriate.
This description must be consistent with exact solutions of
Eq.~(\ref{deteq}) for a large number of particles $N$.

On the single particle level, a statistical description means to replace the
exact, fine-grained force contained in Eq.~(\ref{deteq}) by its smoothed
coarse-grained average force.
The fine-grained aspect of the particle motion is then modeled by an
additional fluctuating force $\boldsymbol{F}_{{\rm L}}$.
As pointed out by Jowett~\cite{jowett}, this concept constitutes ``an
attempt to describe the effects of the neglected microscopic degrees of
freedom''.
In order not to introduce a systematic error into the statistical description
of the $N$-particle ensemble, this force must vanish on the ensemble average
\begin{equation}\label{avgamma}
\< \boldsymbol{F}_{{\rm L}} \> = 0\,.
\end{equation}
In a statistical description, we must not conceive
$\boldsymbol{F}_{{\rm L}}(\boldsymbol{v},t)$
to be an ordinary vector function but a quantity that has only
statistically defined properties.
Fluctuating forces of this nature are usually referred to as
``Langevin forces''~\cite{langevin}.

\begin{figure}
\begin{center}
\epsfig{file=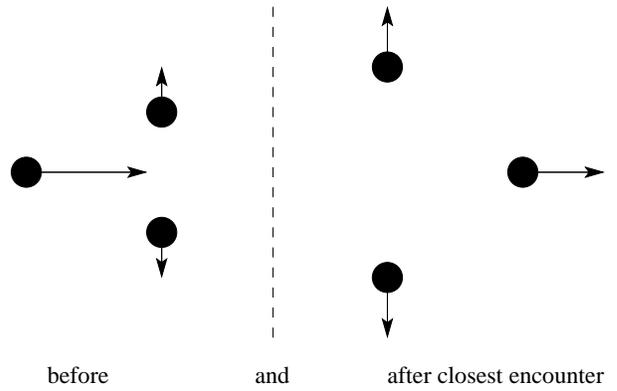}
\vspace{3mm}
\caption{Sketch of the mechanism of dynamical friction for
repelling forces between particles.}
\label{fig:dynfric}
\end{center}
\end{figure}

In performing the transition from an ``exact'' fine-grained description
of the evolution of $\rho(t)$ according to Eq.~(\ref{deteq}) to a
statistical description of this evolution, not only
the fluctuating Langevin force
$\boldsymbol{F}_{{\rm L}}(\boldsymbol{v},t)$ but
also a force that is referred to as ``dynamical friction'' force
$\boldsymbol{F}_{{\rm fr}}(\boldsymbol{v},t)$ must be introduced.
For repelling forces, the mechanism of dynamical friction is sketched in
Fig.~\ref{fig:dynfric}.
We observe that the deceleration of the leftmost particle in horizontal
direction before its closest encounter with the other particles is greater
than its acceleration afterwards.
This means that a net deceleration, hence a friction occurs.
As is easily verified, the same is true for attracting forces.

Owing to the statistical description of the $N$-particle ensemble,
the self-field appears now as a smooth function of $\boldsymbol{x}$ and $t$
that is equivalent to an external force field.
The stochastic counterpart of the deterministic single particle equation
of motion (\ref{deteq}) can now be written as
\begin{equation}\label{stocheq}
m\frac{d^{2}}{dt^{2}} \boldsymbol{x} - \boldsymbol{F}_{{\rm ext}} -
q\boldsymbol{E}_{{\rm sc}}^{{\rm sm}} - \boldsymbol{F}_{{\rm fr}} =
\boldsymbol{F}_{{\rm L}}\,,
\end{equation}
containing the smooth part of the self-force
$\boldsymbol{E}_{{\rm sc}}^{{\rm sm}}(\boldsymbol{x},t)$,
the dynamical friction force
$\boldsymbol{F}_{{\rm fr}}(\boldsymbol{v},t)$, and
the fluctuating Langevin force
$\boldsymbol{F}_{{\rm L}}(\boldsymbol{v},t)$.
As usual, we made the reasonable assumption that the stochastic effects in our
statistical description are independent of the ``external'' force functions
$\boldsymbol{F}_{{\rm ext}}(\boldsymbol{x},t)$ and
$q\boldsymbol{E}_{{\rm sc}}^{{\rm sm}}(\boldsymbol{x},t)$.
This means that the Langevin force $\boldsymbol{F}_{{\rm L}}$
as well as the friction force $\boldsymbol{F}_{{\rm fr}}$ do not
depend on the position $\boldsymbol{x}$ in real space.

Each particle encounters a specific realization of the Langevin force
$\boldsymbol{F}_{{\rm L}}(\boldsymbol{v},t)$.
These forces are defined by their statistical properties only,
a direct integration of Eq.~(\ref{stocheq}) is thus not possible.
On the other hand, a deterministic equation of motion for the phase
space probability density $f(\boldsymbol{x},\boldsymbol{v},t)$
can be derived on the basis of Eq.~(\ref{stocheq}).
This topic will be the subject of the next section.

The statistical mechanics equation (\ref{stocheq}) is supposed to
provide an equivalent description of the dynamics of an $N$
particle ensemble as the ``exact mechanics'' equation (\ref{deteq}).
Therefore, the magnitudes of Langevin and friction forces contained
in Eq.~(\ref{stocheq}) are completely determined by the force
fluctuations that follow from the actual charge granularity,
as described by Eqs.~(\ref{deteq}) and (\ref{distri}).
This is no longer true if Eq.~(\ref{stocheq}) is used as the
analytical basis to interpret results of computer
simulations of $N$ particle ensembles.
If we model the impact of the generally limited accuracy of numerical
methods by Langevin and friction force terms that act on the
simulation particles {\em in addition\/} to ``true'' forces
experienced by the ``real'' beam, the magnitude of the stochastic forces
depends on the specific nature of the simulation code.
In other words, the results of beam simulations must be regarded
as solutions of Eq.~(\ref{stocheq}) with the magnitudes of
$\boldsymbol{F}_{{\rm fr}}$ and $\boldsymbol{F}_{{\rm L}}$ being
determined by the particular scheme of simulation,
defined by the time step width of the numerical integration,
the number of simulation particles, the computer's word length,
and others~\cite{haber}.

In recent studies~\cite{flekk}, the close relation between results
of molecular dynamics (MD) simulations and coarse-grained
dissipative particle dynamics has been worked out.
Similarly, an approach based on the Fokker-Planck equation
in order to explain numerical emittance growth effects
observed in particle-in-cell (PIC) simulations has been presented
earlier~\cite{struck-pa,struck}.
In the following sections, we will review and further extend this analysis.
\section{Fokker-Planck Equation}\label{sec:fokker-planck}
We define $\boldsymbol{q} \equiv (\boldsymbol{x},\boldsymbol{v})$ as the
position vector in the $6$-dimensional $\mu$-phase space.
If the function $f(\boldsymbol{q}, t)$ represents a normalized phase
space probability density, $f\,d\boldsymbol{q}$ provides us with the
probability of finding a particle inside a volume $d\boldsymbol{q}$
around the phase space point $\boldsymbol{q}$ at time $t$.
In these terms, the generalization of Eq.~(\ref{stocheq}) can be written as
\begin{equation}\label{stocheq2}
\dot{q}_{i} = K_{i}(\boldsymbol{q},t) +
\Gamma_{i}(\boldsymbol{q},t)\;,\qquad i=1,\ldots,6\,.
\end{equation}
with smooth functions $K_{i}(\boldsymbol{q},t)$ and the random variables
$\Gamma_{i}(\boldsymbol{q},t)$ vanishing on the ensemble average.
We now assume the random variables $\Gamma_{i}(\boldsymbol{q},t)$
to be Gaussian-distributed and their time correlation
proportional to the $\delta$-function
\begin{equation}\label{corre}
\< \Gamma_{i}(\boldsymbol{q},t)\;\vphantom{F^{L^{L}}}
\Gamma_{j}(\boldsymbol{q},t') \> =
2 Q_{ij}(\boldsymbol{q},t)\,\delta(t-t')\,.
\end{equation}
Under these conditions, the Kramers-Moyal
expansion for
$\partial f(\boldsymbol{q},t) / \partial t$
terminates after the second term~\cite{kramers,moyal,risken}.
The expansion with only the first and second term is usually called
Fokker-Planck equation
\begin{equation}\label{fp1}
\frac{\partial f}{\partial t} = \boldsymbol{L}_{{\rm FP}} f
\end{equation}
with the Fokker-Planck operator $\boldsymbol{L}_{{\rm FP}}$ given by
\begin{displaymath}
\boldsymbol{L}_{{\rm FP}} =
-\sum_{i=1}^{6}\frac{\partial}{\partial q_{i}} K_{i}(\boldsymbol{q},t) +
\sum_{i,j=1}^{6}\frac{\partial^{2}}{\partial q_{i}\partial q_{j}}
Q_{ij}(\boldsymbol{q},t)
\end{displaymath}
We observe that the coefficients $Q_{ij}$ are determined by the
amplitude of the $\delta$-correlated noise functions
$\Gamma_{i}$ according to (\ref{corre}), whereas
the $K_{i}$ are defined by Eq.~(\ref{stocheq2}).
Consequently, Eq.~(\ref{fp1}) represents the deterministic equation of
motion for the probability density $f(\boldsymbol{q},t)$.
It is uniquely determined by the coupled set of Langevin equations
(\ref{stocheq2}) provided that (\ref{corre}) holds.

In terms of the special Langevin equation (\ref{stocheq}), the
Fokker-Planck operator $\boldsymbol{L}_{{\rm FP}}$ reduces to
\begin{equation}\label{fp2}
\boldsymbol{L}_{{\rm FP}} = \sum_{i=1}^{3}\left[
-\frac{\partial}{\partial x_{i}} v_{i} - \frac{1}{m}
\frac{\partial}{\partial v_{i}} F_{{\rm tot},i} +
\frac{\partial^{2}}{\partial v_{i}^{2}} D_{ii}\right]
\end{equation}
with $F_{{\rm tot},i}$ defined as the sum of all non-Langevin forces
\begin{displaymath}
F_{{\rm tot},i}(\boldsymbol{x},\boldsymbol{v},t) =
F_{{\rm ext},i}(\boldsymbol{x},t) +
q E_{{\rm sc},i}^{{\rm sm}}(\boldsymbol{x},t) +
F_{{\rm fr},i}(v_{i},t)\,,
\end{displaymath}
and the diffusion coefficients $D_{ii}$ defined by
\begin{equation}\label{corre2}
\< F_{{\rm L},i}(v_{i},t)\;\vphantom{F^{L^{L}}}
F_{{\rm L},j}(v_{j},t') \> =
2m^{2} D_{ii}(v_{i},t)\delta_{ij}\,\delta(t-t')\,.
\end{equation}
The off-diagonal terms of the diffusion matrix $D_{ij}$ vanish
since the Langevin forces in Eq.~(\ref{stocheq}) are not correlated for
different degrees of freedom.
We further note that the friction forces $F_{{\rm fr},i}$
must always be decelerating.
This means that $F_{{\rm fr},i}$ must change sign if
$v_{i}$ does, hence must be an odd function of $v_{i}$.
With regard to Eq.~(\ref{corre2}), it follows that the diffusion
coefficients of Eq.~(\ref{fp2}) must be even functions of the $v_{i}$
\begin{equation}\label{negvi}
F_{{\rm fr},i}(v_{i}) = -F_{{\rm fr},i}(-v_{i})\quad,\quad
D_{ii}(v_{i}) = D_{ii}(-v_{i})\,.
\end{equation}
A Fokker-Planck equation that describes the evolution of the probability
density $f$ appertaining to the stochastic motion of particles in external
force fields if often referred to as Kramers equation.
As will be shown in the next section investigating equilibrium solutions of
Eq.~(\ref{fp1}) with the Fokker-Planck operator (\ref{fp2}), the diffusion
coefficients $D_{ii}(v_{i},t)$ are uniquely determined by the
friction forces $F_{{\rm fr},i}(v_{i},t)$.
\section{Fokker-Planck Coefficients under Time Reversal}\label{sec:time-rev}
If we perform a transformation that reverses the direction of time flow,
the positions $x_{i}$ and hence all quantities that only depend on the
positions do not change sign.
In contrast, the velocities $v_{i}$ do change sign, which means that
quantities depending on the $v_{i}$ may change sign under time reversal.
We may thus separate the components of the Fokker-Planck operator
(\ref{fp2}) with respect to their behavior under time reversal
\begin{displaymath}
\boldsymbol{L}_{{\rm FP}} = \boldsymbol{L}_{{\rm rev}} +
\boldsymbol{L}_{{\rm ir}}\,.
\end{displaymath}
The ``reversible'' operator $\boldsymbol{L}_{{\rm rev}}$ is defined to
consist of those components of (\ref{fp2}) that change sign under time reversal
\begin{equation}\label{fp-rev}
\boldsymbol{L}_{{\rm rev}} = \sum_{i=1}^{3}\left[
-\frac{\partial}{\partial x_{i}} v_{i} - \frac{1}{m}
\frac{\partial}{\partial v_{i}} \left( F_{{\rm ext},i} +
q E_{{\rm sc},i}^{{\rm sm}}\right)\right]\,.
\end{equation}
The smooth self-field $\boldsymbol{E}_{{\rm sc}}^{{\rm sm}}$
is obtained from the real space projection of the probability density
$f(\boldsymbol{q},t)$ via Poisson's equation.
The components that do not change sign constitute
$\boldsymbol{L}_{{\rm ir}}$
\begin{equation}\label{fp-ir}
\boldsymbol{L}_{{\rm ir}} = \sum_{i=1}^{3}
\frac{\partial}{\partial v_{i}}\left[
-\frac{F_{{\rm fr},i}(v_{i},t)}{m} +
\frac{\partial}{\partial v_{i}} D_{ii}(v_{i},t)\right]\,.
\end{equation}
Here we made use of (\ref{negvi}), which states that under time reversal
$F_{{\rm fr},i}$ changes sign, whereas $D_{ii}$ does not change sign.
The external forces $F_{{\rm ext},i}$ have been assumed to be not
velocity dependent.

Since $\partial f / \partial t$ changes sign on time reversal, a
Fokker-Planck equation with only $\boldsymbol{L}_{{\rm rev}}$
remains unchanged if the direction of time flow is reversed.
It therefore describes the reversible transformation of the probability
density function $f(\boldsymbol{x},\boldsymbol{v},t)$.
This means that earlier states are fully restored if a reversed time
integration of Eq.~(\ref{fp1}) with
$\boldsymbol{L}_{\rm FP}\equiv\boldsymbol{L}_{\rm rev}$
is carried out --- just like a movie
that is reversed at some instant of time $t_{0}$.
Correspondingly, $\boldsymbol{L}_{{\rm ir}}$ describes exactly those
effects that do {\em not\/} depend on the direction of the time flow.
In other words, it describes the irreversible aspects of the particle motion.
With $\boldsymbol{L}_{{\rm ir}} = 0$,
Eq.~(\ref{fp1}) is commonly referred to as Vlasov equation.
\section{Equilibrium Distributions in Autonomous Systems}\label{sec:autosyst}
If the external force $\boldsymbol{F}_{{\rm ext}}(\boldsymbol{x})$
contained in Eq.~(\ref{fp2}) is not explicitly time dependent, a
stationary solution $\boldsymbol{L}_{{\rm FP}} f_{{\rm st}} = 0$ may exist.
If it exists, it can always be written in the form
\begin{equation}\label{f-stat}
f_{{\rm st}}(\boldsymbol{x},\boldsymbol{v}) =
g_{0}^{-1}\exp\left\{-\phi_{{\rm st}}
(\boldsymbol{x},\boldsymbol{v})\right\}\,,
\end{equation}
with $g_{0} = \int\exp\left\{-\phi_{{\rm st}}(\boldsymbol{x},\boldsymbol{v})
\right\}d\boldsymbol{x}d\boldsymbol{v}$ the normalization factor.
We may define the irreversible probability current $S^{{\rm ir}}_{v_{i}}$
flowing into the $v_{i}$-direction in phase space as
\begin{equation}\label{irpc}
\boldsymbol{L}_{{\rm ir},i} f = -\frac{\partial}{\partial v_{i}}
S^{{\rm ir}}_{v_{i}}\,.
\end{equation}
Obviously, all irreversible currents must
vanish for $f=f_{{\rm st}}$ to be stationary.
With $\boldsymbol{L}_{{\rm ir},i}$ given by (\ref{fp-ir}),
this means, explicitly,
\begin{equation}\label{fluc-diss}
\frac{F_{{\rm fr},i}(v_{i})}{m} = \frac{\partial D_{ii}(v_{i})}
{\partial v_{i}} - D_{ii}(v_{i})
\frac{\partial\phi_{{\rm st}}(\boldsymbol{x},\boldsymbol{v})}
{\partial v_{i}}\,.
\end{equation}
Eq.~(\ref{fluc-diss}) states that for given $\phi_{{\rm st}}$, the
diffusion function $D_{ii}(v_{i})$ is uniquely determined by the friction
force function $F_{{\rm fr},i}$ --- and vice versa.
This mutual dependency of the diffusion effects --- driving a system away
from its steady state --- and damping effects that cause the decay of these
deviations makes up the physical substance of ``fluctuation-dissipation
theorems''.

In agreement with Eq.~(\ref{negvi}), we express the friction force
function $F_{{\rm fr},i}$, and the diffusion function $D_{ii}$
as odd and even power series in $v_{i}$, respectively
\begin{equation}\label{powerseries}
F_{{\rm fr},i}(v_{i})=-m\sum_{k=0}^{\infty}a_{k}v_{i}^{2k+1}\;,\;
D_{ii}(v_{i}) = \sum_{k=0}^{\infty}b_{k}v_{i}^{2k}\,.
\end{equation}
Here we assumed the coefficients $a_{k}, b_{k}$ not to depend on
$\boldsymbol{x}$ and the degree of freedom $i$ --- in agreement with the
precondition that the stochastic effects are not influenced by the external
forces.

With (\ref{powerseries}), we find that Eq.~(\ref{fluc-diss}) can only be
fulfilled if $\phi_{{\rm st}}$ is a quadratic function of the $v_{i}$.
Therefore, $\phi_{{\rm st}}$ may always be separated as
\begin{equation}\label{f-stat2}
\phi_{{\rm st}}(\boldsymbol{x},\boldsymbol{v}) =
\psi_{{\rm st}}(\boldsymbol{x}) +
\sum_{i=1}^{3}\frac{v_{i}^{2}}{2\<v_{i}^{2}\>}\,,
\end{equation}
the angle brackets denoting the respective averages over the
phase space density function: $\< a\>=\int a f d\tau$.
The quantity $\<v_{i}^{2}\>$ thus embodies the ensemble average of
the squares of all particle velocities, also referred to as the
second moment of the velocity $v_{i}$ of the equilibrium
distribution $f_{{\rm st}}$.
In a state of equilibrium these moments must agree for each degree
of freedom, hence can be identified with the
equilibrium temperature $T_{{\rm eq}}$ according to
\begin{equation}\label{T-equil}
kT_{{\rm eq}} = m\< v_{i}^{2} \>\,,\, i=1,2,3\,,
\end{equation}
with $k$ denoting Boltzmann's constant.

Inserting (\ref{f-stat2}) into the Fokker-Planck equation
(\ref{fp1},~\ref{fp2}), the generalized potential
$\psi_{{\rm st}}(\boldsymbol{x})$ follows from
\begin{equation}\label{genpot}
\nabla\psi_{{\rm st}}(\boldsymbol{x}) =
-\frac{1}{kT_{{\rm eq}}} \left( \vphantom{F^{x^{x}}}
\boldsymbol{F}_{{\rm ext}}(\boldsymbol{x}) +
q \boldsymbol{E}_{{\rm sc}}^{{\rm sm}}(\boldsymbol{x})\right)\,,
\end{equation}
In final form, the equilibrium probability density of the Fokker-Planck
equation (\ref{fp1}) reads
\begin{equation}\label{f-stat3}
f_{{\rm st}} = g_{0}^{-1}\exp\left\{ -\psi_{{\rm st}}(\boldsymbol{x})
\right\} \exp\left\{-\sum_{i=1}^{3}
\frac{mv_{i}^{2}}{2kT_{{\rm eq}}}\right\}\,.
\end{equation}
We summarize that the equilibrium distribution (\ref{f-stat3})
follows directly from the assumption that the stochastic component of
the particle motion is caused by Gaussian-distributed Langevin forces
with a time correlation function proportional to the $\delta$-function,
regardless of the dependency of the friction and the diffusion
coefficients on the $v_{i}$.
For a given temperature $T_{{\rm eq}}$,
the spatial probability function following from $\psi_{{\rm st}}$
is uniquely determined by the external force
$\boldsymbol{F}_{{\rm ext}}$, and the stationary self-field
$\boldsymbol{E}_{{\rm sc}}^{{\rm sm}}$.
Together with the unique velocity distribution, the entire phase space
probability density function is uniquely determined, which means that
no other equilibrium distribution of (\ref{fp1}) exists --- in contrast to
Vlasov systems where friction as well as diffusion effects vanish.
If the external force function $F_{{\rm ext}}(\boldsymbol{x})$
does allows for an equilibrium, and if the friction is not negligible,
arbitrary non-equilibrium density functions always settle down to a
unique equilibrium.
This is what we observe in long-term simulations of charged
particle beams~\cite{reiser,lund}.
Regardless of our initial phase space filling, we always end up with a
Gaussian velocity distribution if no resonance effects are involved.
\section{Boltzmann Entropy Growth}\label{sec:boltz}
As a simple ansatz,
we truncate the power series (\ref{powerseries}) for the friction force
$F_{{\rm fr},i}(v_{i})$ after the cubic term in $v_{i}$
\begin{equation}\label{powerseries1}
\frac{1}{m}F_{{\rm fr},i}(v_{i}) = -\gamma_{1}v_{i}-\gamma_{3}
\frac{m}{kT_{{\rm eq}}}v_{i}^{3}\,.
\end{equation}
With (\ref{fluc-diss}) and (\ref{f-stat2}) we then immediately obtain
the diffusion coefficient
\begin{equation}\label{powerseries2}
D_{ii}(v_{i}) = \gamma_{1}\frac{kT_{{\rm eq}}}{m} + \gamma_{3}\left(
v_{i}^{2}+2\frac{kT_{{\rm eq}}}{m}\right).
\end{equation}
In order to quantify the impact of friction and diffusion on
the phase space probability density function
$f(\boldsymbol{x},\boldsymbol{v},t)$ we now define the negative
Boltzmann entropy $S(t)$ as~\cite{liboff,struck}
\begin{equation}\label{entdef}
S(t)=-k\int f\,\ln f \,d\boldsymbol{x}d\boldsymbol{v}\,.
\end{equation}
We easily convince ourselves that $S(t)$ remains unchanged if only
the reversible part (\ref{fp-rev}) of the Fokker-Planck operator
$\boldsymbol{L}_{{\rm FP}}$ drives the time evolution of $f$
\begin{displaymath}
\frac{\partial f}{\partial t} = \boldsymbol{L}_{{\rm rev}}f\;\;
\Longrightarrow\;\; S(t)={\rm const.}
\end{displaymath}
If a phase space density $f$ does not represent an equilibrium state,
the irreversible part (\ref{fp-ir}) of $\boldsymbol{L}_{{\rm FP}}$
must therefore account for a non-constant $S(t)$.
We approximate an arbitrary non-equilibrium phase space probability
density distribution $f(\boldsymbol{x},\boldsymbol{v},t)$ by
\begin{equation}\label{non-equil}
f(\boldsymbol{x},\boldsymbol{v},t)=
g(\boldsymbol{x},t)\,\exp\left\{ -\sum_{i=1}^{3}
\frac{m{\left(v_{i}^{{\rm inc}}\right)}^{2}}{2 k T_{i}}\right\}\,,
\end{equation}
with $g(\boldsymbol{x},t)$ the non-equilibrium real space probability
density, $T_{i}(t)$ the non-equilibrium ``temperature'' appertaining to
the $i$-th degree of freedom, and $v_{i}^{{\rm inc}}$ the incoherent
contribution to the total particle velocity $v_{i}$
\begin{displaymath}
v_{i}^{{\rm inc}} = v_{i} - x_{i}\frac{\< x_{i}v_{i}\>}{\< x_{i}^{2}\>}\,.
\end{displaymath}
In energy units, the temperature $kT_{i}(t)=m\< v_{i}^{2}\>_{{\rm inc}}$
is defined as the incoherent kinetic beam energy.
If a phase space distribution $f$ is in non-equilibrium state,
the total kinetic energy of a beam particle consists of both,
a coherent as well as an incoherent part.
The temperature $kT_{i}(t)$ is thus obtained by subtracting
the coherent kinetic energy from the total kinetic beam energy.
$kT_{i}(t)$ then evaluates to
\begin{equation}\label{T-def}
kT_{i}(t) = m \frac{\varepsilon_{i}^{2}}{\< x_{i}^{2}\>}\;,\quad
\varepsilon_{i}^{2} = \< x_{i}^{2}\> \< v_{i}^{2}\> - \< x_{i}v_{i}\>^{2}\,,
\end{equation}
with $\varepsilon_{i}(t)$ denoting the root mean square (RMS) emittance
of the particle ensemble in the $i$-th degree of freedom.
With the non-equilibrium density function (\ref{non-equil}), and
$\boldsymbol{L}_{{\rm ir}}$ according to (\ref{fp-ir}), we obtain for
the time derivative of the entropy (\ref{entdef})~\cite{struck}
\begin{equation}\label{entmark}
\frac{d S}{d t} = k \sum_{i=1}^{3} \left(
\< \frac{\partial F_{{\rm fr},i}/m}{\partial v_{i}} \> +
\frac{m}{kT_{i}} \< D_{ii} \> \right)\,.
\end{equation}
Inserting our approximations for the friction force
(\ref{powerseries1}) and the diffusion coefficient
(\ref{powerseries2}), this expression simplifies to
\begin{equation}\label{entmark1}
\frac{d S}{d t} = k\gamma
\sum_{i=1}^{3} \left( \frac{T_{{\rm eq}}}{T_{i}} - 1 \right)\;,\quad
\gamma = \gamma_{1} + 3\gamma_{3} > 0\,.
\end{equation}
We note that the cubic term of the friction force ansatz
(\ref{powerseries1}) does not modify the form of Eq.~(\ref{entmark1}).

The equilibrium temperature $T_{\rm eq}$ has been defined in
Eq.~(\ref{T-equil}) for autonomous systems.
Before applying Eq.~(\ref{entmark1}) to non-autonomous
systems, we must discuss how to define appropriately the
``equilibrium temperature'' $T_{\rm eq}$ in these systems.
This will be the subject of the following section.
\section{Equilibrium Temperature in Non-Autonomous Systems}\label{sec:equil}
For equilibrium distributions in autonomous systems as
discussed in Sec.~\ref{sec:autosyst}, the equilibrium temperature
$T_{{\rm eq}}$ contained in (\ref{entmark1}) is a constant of motion.
For real focusing systems, i.e.\ non-autonomous systems with
the external force $\boldsymbol{F}_{{\rm ext}}(\boldsymbol{x},t)$ being
explicitly time dependent in the beam system, such a constant does not exist.
Under these circumstances, an equilibrium temperature must be defined
analogically as the mean temperature averaged over a correlation time
interval $\delta\tau$.
For $t\ge\delta\tau$ this means
\begin{equation}\label{T-equil1}
T_{{\rm eq}}(t;\delta\tau)=\frac{1}{3\delta\tau}
\int\limits_{t-\delta\tau}^{t} \sum_{i=1}^{3} T_{i}(t') \, dt'\,.
\end{equation}
The length of the time interval $\delta\tau$ must depend on the
amplitude of the Langevin forces acting within our system, i.e.\
on the effective friction coefficient $\gamma$.
For $\gamma\to 0$, hence for reversible systems,
the memory on earlier states is never lost.
Therefore the average temperature depends on the whole time
interval elapsed since launching of the beam
\begin{displaymath}
T_{{\rm eq}}(t;\delta\tau\!\!\to\!\infty)=\frac{1}{3t}
\int\limits_{0}^{t} \sum_{i=1}^{3} T_{i}(t') \, dt'\,.
\end{displaymath}
On the other hand, if $\gamma$ is very large, the memory on earlier
states is rapidly lost, which means that $\delta\tau\to 0$.
The equilibrium temperature is then given by the instantaneous
temperature $T(t)$, namely the arithmetic mean of the temperatures $T_{i}(t)$
\begin{displaymath}
T_{{\rm eq}}(t;\delta\tau\!\!\to\!0)=\frac{1}{3}\sum_{i=1}^{3} T_{i}(t)\,.
\end{displaymath}
In real systems --- as well as in numerical simulations that take
into account the space charge forces --- $\gamma$ is usually small but finite.
As a consequence, the equilibrium temperature definition
(\ref{T-equil1}) must be used in our analytical description.
Inserting (\ref{T-equil1}) into (\ref{entmark1}), the entropy
change over the time span $\delta\tau$ follows by integration
\begin{equation}\label{entgro}
S(t)-S(t-\delta\tau) = 3k\gamma\,\delta\tau\,A_{\delta\tau}(t)\,,
\end{equation}
with the dimensionless temperature anisotropy coefficient
$A_{\delta\tau}(t)$ defined as
\begin{equation}\label{aniso}
A_{\delta\tau}(t)=\left[\frac{1}{{(3\delta\tau)}^{2}}\!\!\!
\int\limits_{t-\delta\tau}^{t}\sum_{i=1}^{3}\frac{\< x_{i}^{2}\>}
{\varepsilon_{i}^{2}}dt'\!\!\!\int\limits_{t-\delta\tau}^{t}
\sum_{j=1}^{3}\frac{\varepsilon_{j}^{2}}{\< x_{j}^{2}\>}dt'\right] - 1\,.
\end{equation}
The total entropy change at multiples of the correlation time interval
$t=M\delta\tau$ is then given by the sum over all elementary intervals
\begin{equation}\label{entgrototal}
S(t)-S(0)=3k\,\frac{\gamma t}{M}\,
\sum_{m=1}^{M}A_{\delta\tau}(m\,\delta\tau)\,.
\end{equation}
Eq.~(\ref{entgrototal}) states that the entropy growth a particle
ensemble experiences is determined by both, the friction coefficient
$\gamma$, and the average temperature imbalance according
to Eq.~(\ref{aniso}).
It represents what we expect for a Markov process: the total entropy
increase is given by a sum of elementary increases that do not depend
on each other.
The temperature imbalance is determined by the optical properties
of the focusing lattice, the matching conditions of the beam, and
the finite correlation time $\delta\tau>0$ that measures the
typical duration of an elementary scattering event.
$\delta\tau$ thus defines the time span over which the
instantaneous temperatures $T_{i}(t)$ must be averaged in
order to obtain the reference temperature (\ref{T-equil1}).

The actual value of $\gamma$ can be estimated for charged
particle beams by averaging the elementary process of a binary
Coulomb scattering event over the impact parameters, and
subsequently over the beam's velocity distribution~\cite{jansen}.

For a beam propagation that is being influenced by a large number of
internal scattering events, irreversibility just means that the
time-reversed evolution is highly {\em improbable}.
This perception of irreversible behavior of a dynamical system cannot
be applied directly to numerical noise effects in computer simulations.
For this case, a more appropriate interpretation of the entropy
has been given by Shannon~\cite{shannon}, in the context of his
founding of ``information theory''.
Within this framework, the entropy measures the amount of missing
information about the state of the system in question.
The entropy growth thus quantifies the amount of information
an observer looses during the system's time evolution.
In case of our simulations, the necessarily limited accuracy
of numerical methods accounts for the mechanism that causes a loss
of information.
We can therefore no longer regard $\gamma$ to represent a ``real''
physical process if we make use of Eq.~(\ref{entgrototal})
in order to interpret beam simulation results.
Unfortunately, it appears to be impossible to analytically
estimate the ``computer noise'' induced $\gamma$ in so far as
it depends on the particular realization of the simulation.
The actual value of $\gamma$ must therefore be extracted from the
simulation data itself by comparing the obtained emittance growth
factors for different average temperature imbalances.
To this end, the relation between entropy change (\ref{entgrototal})
and irreversible emittance growth must be established.
This will be worked out in the following section.
Afterwards, the discussion to what extend Eq.~(\ref{entgrototal})
can also be used to interpret computer simulation results
will be presented in Sec.~\ref{sec:simu}.
\section{Emittance Growth associated with
$\boldsymbol{L}_{\rm ir}$}\label{sec:fpegr}
A second order moment analysis of Eq.~(\ref{fp1}) yields the
following set of coupled moment equations~\cite{struck-pa}
for each phase space plane $i = 1,2,3$
\begin{eqnarray*}
\frac{d}{d t} \<x_{i}^{2}\> &\!=\!& 2 \<x_{i} v_{i}\>\\
\frac{d}{d t} \<x_{i} v_{i}\> &\!=\!& \<v_{i}^{2}\> + \frac{1}{m}\Big[
\<x_{i} F_{{\rm ext},i}\>\!+\!q
\<x_{i} E^{{\rm sm}}_{{\rm sc},i}\>\!+\!
\<x_{i} F_{{\rm fr},i}\>\Big]\\
\frac{d}{d t}\<v_{i}^{2}\> &\!=\!& \frac{2}{m}\Big[
\<v_{i} F_{{\rm ext},i}\>\!+\!q
\<v_{i} E^{{\rm sm}}_{{\rm sc},i}\>\!+\!
\<v_{i} F_{{\rm fr},i}\>\!\Big] \!+\! 2\< D_{ii} \>
\end{eqnarray*}
Calculating the time derivative of the RMS emittance (\ref{T-def}),
and inserting the above derivatives of the second moments,
we find that three distinct sources for the RMS emittance change
can be distinguished
\begin{equation} \label{depsdt}
\frac{d}{dt}\varepsilon_i^2(t) =
\left.\frac{d}{dt}\varepsilon_i^2(t)\right|_{{\rm ext}} +
\left.\frac{d}{dt}\varepsilon_i^2(t)\right|_{{\rm sc}} +
\left.\frac{d}{dt}\varepsilon_i^2(t)\right|_{{\rm ir}}\,,
\end{equation}
namely the external field contribution, the contribution
related to the smooth space charge fields and the contribution
due to the Langevin forces described by the irreversible part
(\ref{fp-ir}) of the Fokker-Planck operator.

Collecting together the terms containing the external force
$\boldsymbol{F}_{{\rm ext}}$, we find that its contribution
to the change of the RMS emittance vanishes if it is a linear
function of the spatial coordinates
\begin{eqnarray*}
\left.\frac{m}{2}
\frac{d}{dt}\varepsilon_i^2(t)\right|_{{\rm ext}} \!\!\! & = &
\<x_i^2\> \<v_i F_{{\rm ext},i} \> - \<x_i v_i\>
\<x_i F_{{\rm ext},i}\> \\
& = & 0\quad \Longleftrightarrow \quad  F_{{\rm ext},i} \propto x_{i}\,.
\end{eqnarray*}
This applies to all our computer simulations
addressed in Sec.~\ref{sec:simu}.

The terms containing the smooth space charge field
$\boldsymbol{E}^{{\rm sm}}_{{\rm sc}}$ sum up to
\begin{displaymath}
\left.\frac{m}{2}
\frac{d}{dt}\varepsilon_{i}^{2}(t)\right|_{{\rm sc}} =
q\left[ \<x_{i}^{2}\>\<v_{i} E^{{\rm sm}}_{{\rm sc},i}\>
- \<x_{i} v_{i}\>\<x_{i} E^{{\rm sm}}_{{\rm sc},i}\>\right]\,.
\end{displaymath}
Writing this equation for all three spatial degrees of freedom,
the electric field terms together form the physical quantity
of ``free field energy'', i.e.\ the difference of the {\em actual\/}
field energy $W$ and the field energy $W_{{\rm u}}$ of the
equivalent {\em uniform\/} charge density~\cite{wangler,host,pacc1}
\begin{equation}\label{wangl}
\left.\sum_{i=1}^{3}\frac{1}{\<x_{i}^{2}\>}
\frac{d}{d t}\varepsilon_{i}^{2}(t)\right|_{{\rm sc}} +
\frac{2}{mN}\frac{d}{d t} \left( W - W_{{\rm u}} \right) = 0\,.
\end{equation}
In Sec.~\ref{sec:simu}, we will show by numerical simulation
that the exchange of RMS emittance and ``free field energy''
is indeed a reversible process.

The third contribution to the change of the RMS emittance emerges
from the irreversible Fokker-Planck operator~(\ref{fp-ir})
\begin{equation}\label{depsdtFP}
\left.\frac{m}{2}
\frac{d}{dt}\varepsilon_{i}^{2}(t)\right|_{{\rm ir}}\!\! =
\<x_{i}^{2}\> \<v_{i} F_{i} \> - \<x_{i} v_{i}\> \<x_{i} F_{i}\> +
m\<x_{i}^{2}\> \< D_{ii}\>.
\end{equation}
We restrict ourselves to a friction force (\ref{powerseries1})
linear in the $v_{i}$.
The related diffusion coefficient $D_{ii}$ then follows as
\begin{equation}\label{o-u-coeff}
\frac{1}{m}F_{{\rm fr},i} = -\gamma_{1}v_{i}\;,\qquad
D_{ii} = \gamma_{1}\frac{kT_{{\rm eq}}}{m}\,.
\end{equation}
This ansatz corresponds to Stokes's friction law in classical mechanics.
It applies if the Langevin forces are small in comparison
to all other forces relevant for the dynamics of the system.
Inserting (\ref{o-u-coeff}) into Eq.~(\ref{depsdtFP}) we obtain
\begin{equation}\label{dlnepsdt2}
\left.\frac{d}{dt}\ln\varepsilon_{i}(t)\right|_{{\rm ir}} =
\gamma_{1} \,\left(\frac{T_{{\rm eq}}}{T_{i}} - 1 \right)\,.
\end{equation}
Since $\gamma_{1}$ is always positive, the equation states that
the RMS emittance $\varepsilon_{i}(t)$ increases, as long as the
beam temperature $T_{i}$ lies below the equilibrium temperature
$T_{{\rm eq}}$ --- and vice versa.
This is what we expect for a temperature balancing process.
We observe that the right hand sides of Eqs.~(\ref{dlnepsdt2}) and
(\ref{entmark1}) agree for $\gamma=\gamma_{1}$.
This means that the growth of the Boltzmann entropy (\ref{entdef})
is related to the irreversible emittance growth according to
\begin{displaymath}
\left.\frac{1}{k} \frac{dS}{dt} = \frac{d}{dt}\ln
\varepsilon_{x}(t)\varepsilon_{y}(t)\varepsilon_{z}(t)\right|_{{\rm ir}}\,.
\end{displaymath}
With Eq.~(\ref{entgrototal}) and
$\varepsilon=\sqrt[3]{\varepsilon_{x}\varepsilon_{y}\varepsilon_{z}}$,
the related irreversible
growth of the total RMS emittance at $t=M\delta\tau$ is given by
\begin{equation}\label{epsgro}
\left.\ln\frac{\varepsilon(t)}{\varepsilon(0)}\right|_{{\rm ir}} =
\frac{\gamma_{1} t}{M}\sum_{m=1}^{M} A_{\delta\tau}(m\,\delta\tau)\,.
\end{equation}
The validity of Eq.~(\ref{epsgro}) will be verified by numerical
simulations in Sec.~\ref{sec:simutab}.
Beforehand, we demonstrate the emerging of irreversibility due
to ``computer noise'' effects that necessarily accompany
numerical simulations of charged particle beams.
\section{Numerical Simulations}\label{sec:simu}
\subsection{Initial Emittance Change}\label{sec:iniemi}
In this section, we present results of a simulation code that numerically
integrates the reversible equations of motion~(\ref{deteq}), starting
from a sharply known initial phase space filling (\ref{distri}).

As the first example, we simulate the transient effect of ``initial''
emittance change --- a well understood phenomenon in the
realm of charged particle beams~\cite{strklarei,wangler,reiser}.
It occurs if a beam is injected into an ion optical
system in a non-equilibrium state.
As a consequence, the phase space density $f$ changes rapidly
until an average equilibrium state is reached.
As part of this process, the charge density profile adjusts itself
to the external forces, which means that the electrostatic field
energy constituted by the initial charge density is modified.
This process is accompanied by a change of the RMS emittances.
With the emittance definition of Eq.~(\ref{T-def}), the ``exchange''
of emittance and field energy is described quantitatively
by Eq.~(\ref{wangl}).
\begin{figure}
\begin{center}
\epsfig{file=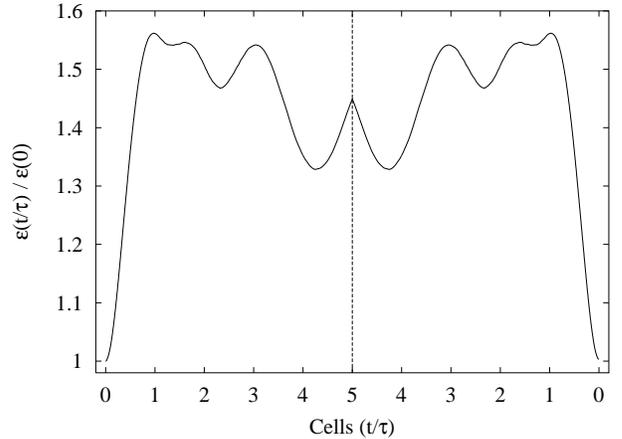,height=86mm,angle=-90}
%\vspace*{3mm}
\caption{Emittance growth factors versus number of cells obtained
for a non-stationary initial phase space density
at $\sigma_0=60^\circ$, $\sigma=15^\circ$, 2500 simulation particles.
The dashed line marks the point of the time reversal after 5~cells.}
\label{fig:5cells}
\end{center}
\end{figure}
We notice that this equation is invariant with respect to
the reversal of time --- in agreement with the fact that
it follows directly from the Vlasov equation
$\partial f/\partial t = \boldsymbol{L}_{{\rm rev}} f$.
The conclusion that the initial charge density adjustment is indeed
a reversible process is verified by the simulation results displayed
in Fig.~\ref{fig:5cells}.
Because of a peaked initial charge density defined by the initial
phase space filling, field energy is released
immediately after launching the beam, followed by a damped
oscillation around the field energy of the equilibrium density profile.
This process is accompanied by a variation of the RMS emittance
according to Eq.~(\ref{wangl}).
Nonetheless, if the direction of the beam transformation is reversed
after $5$ periods, the initial non-equilibrium state is recovered.

\begin{figure}
\begin{center}
\epsfig{file=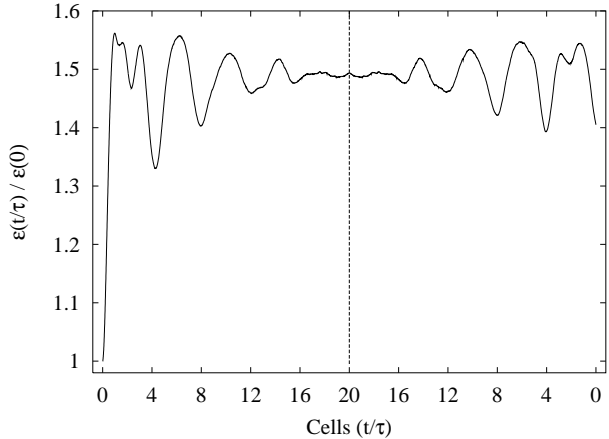,height=86mm,angle=-90}
%\vspace{3mm}
\caption{Emittance growth factors versus number of cells obtained
for a non-stationary initial phase space density
at $\sigma_0=60^\circ$, $\sigma=15^\circ$ per cell, 2500 simulation particles.
The dashed line marks the point of the time reversal after 20~cells.}
\label{fig:20cells}
\end{center}
\end{figure}

This is no longer true if the forward transformation exceeds
a certain amount of periods.
Fig.~\ref{fig:20cells} shows the emittance variations obtained from
the similar simulation as in the previous case, but with the forward and the
subsequent backward transformation now extending over $20$ focusing periods.
Obviously, the RMS emittance does no longer return to its initial value,
but keeps on oscillating around the level associated with the
self-consistent state.
After having been transformed over a certain time span, the
simulated beam has evolved in a way that cannot be reversed anymore.
This behavior can be explained if we interpret the numerical
inaccuracies that inevitably accompany all our simulations to arise
from forces $\boldsymbol{F}_{\rm fr}$ and $\boldsymbol{F}_{\rm L}$
of Eq.~(\ref{stocheq}) that additionally act on the simulation particles.
As outlined in Sec.~\ref{sec:fokker-planck}, these forces induce a
non-vanishing irreversible component of the Fokker-Planck operator
$\boldsymbol{L}_{{\rm ir}}\neq 0$ in the equation of motion for
the phase space density function $f$.
Accordingly, we obtain a gradual loss of ``memory'' of previous states
of the phase space density $f(\boldsymbol{x},\boldsymbol{v},t)$.
This loss of information during the beam transformation
gradually renders all emittance growth effects irreversible,
regardless of their specific nature being reversible or irreversible.
In other words, after having passed $20$ cells, in our particular
simulation example the beam does not ``remember'' anymore that a reversible
emittance growth has taken place right after the start of the simulation.
As the consequence, the specific initial beam state that is
assigned to a lower emittance cannot be recovered.
Instead, the more probable self-consistent state
associated with the increased emittance is kept.
\subsection{Emittance Growth due to Anisotropic Focusing}\label{sec:long-term}
\begin{figure}
\begin{center}
\vspace{-5mm}
\epsfig{file=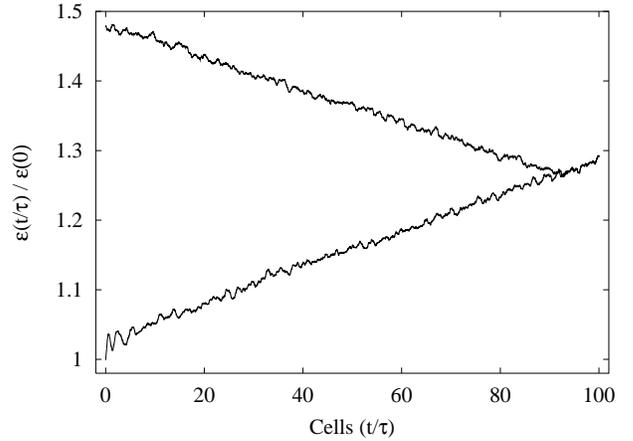,height=86mm,angle=-90}
\vspace{3mm}
\caption{Emittance growth factors versus number of cells obtained
by 3-D simulations of a periodic non-isotropic focusing system
at $\sigma_0=60^\circ$, $\sigma=15^\circ$ per cell,
2000 simulation particles. After 100 cells the time reversal occurs.}
\label{fig:100cells-tr}
\end{center}
\end{figure}

Fig.~\ref{fig:100cells-tr} displays the emittance growth curve obtained
for a simulation of beam transport through a fictitious lattice
that focuses the beam anisotropically in all three spatial directions.
This anisotropic focusing enforces a non-vanishing beam temperature
anisotropy coefficient~(\ref{aniso}) throughout the lattice.
As stated in the previous subsection, numerical simulations
are always accompanied by ``discreteness errors'', which may be
described by additional Langevin and friction force terms in the
single particle equation of motion~(\ref{stocheq}).
With regard to our stochastic description of beams, this means that
both quantities, the finite temperature anisotropy $A_{\delta\tau}$
as well as a non-vanishing effective friction coefficient $\gamma_{1}$
induce a specific irreversible growth rate of the beam emittance
according to Eq.~(\ref{epsgro}).
The amplitudes of the Langevin forces emerging in beam simulations
are larger than the corresponding charge granularity forces
occurring within a real beam.
We thus encounter a larger friction coefficient $\gamma_{1}$
within our simulation procedure than we would expect from an
analytical estimation of $\gamma_{1}$ for a real beam~\cite{jansen}.
Consequently, the numerically obtained emittance growth rate must be
larger than the intrabeam scattering related growth rate for a real beam.
In order to verify that the emittance growth rate obtained in our
simulation of Fig.~\ref{fig:100cells-tr} is indeed caused by the action
of Langevin forces, we again numerically reverse the direction of the
time integration of the single particle equations of motion.
For a small number of focusing periods --- in this particular case
for about 6 periods --- the beam evolution ``behaves'' reversible,
visualized by the ``roll back'' of the emittance curve that
covers exactly the forward transformation graph.
Having exceeded this time span, the beam's evolution in rendered
irreversible, indicated by the sharp change of sign of the slope
of the emittance curve.
This behavior is directly related to the presence of Langevin forces
which, as noted earlier, induce a non-vanishing irreversible part
of the Fokker-Planck operator~(\ref{fp-ir}).
As shown in Sec.~\ref{sec:time-rev}, this operator
describes exactly those aspects of the evolution of a beam dynamical
system that do not change if the direction of time flow is reversed.
This is what we observe in Fig.~\ref{fig:100cells-tr} after
the reversible phase of the back transformation.
The emittance growth rate persisting during the irreversible phase of
the back transformation agrees exactly with the growth rate obtained
along the preceding forward transformation.
\subsection{Scaling Law for the Friction Coefficient}\label{sec:simutab}
We conclude this article by presenting simulation results aimed at
investigating the range of validity of Eq.~(\ref{epsgro}).
As shown in Sec.~\ref{sec:fpegr}, this equation relates the
logarithm of the irreversible emittance growth to the product
of the effective friction coefficient $\gamma_{1}$ and the average
temperature anisotropy the beam experiences between $0$ and $t$.
Accordingly, the simulations comprise examples of beam tracking
through various focusing systems and beam matching conditions,
each of them inducing a specific temperature anisotropy
$A_{\delta\tau}$ along the beam line.
The correlation time parameter $\delta\tau$ contained herein has been
introduced in Eq.~(\ref{T-equil1}) as constituent part of the definition
of the ``equilibrium temperature'' for non-autonomous systems.
For a given friction parameter $\gamma_{1}$, this correlation time
$\delta\tau$ must be adjusted appropriately.
Furthermore, similar simulations are performed with
only the numbers of simulation particles being varied.
We hereby modify the friction coefficient $\gamma_{1}$ in order to
verify Eq.~(\ref{epsgro}) independently for different noise levels.

Unlike the beam entropy, the beam's RMS emittance can be calculated
directly from positions and velocities of all simulation particles.
Unfortunately, as stated by Eq.~(\ref{depsdt}), different mechanisms
that all modify the RMS emittance in the course of the beam
propagation must be distinguished in order to isolate the
noise-related emittance growth effects.
Because of the linear external focusing forces applied throughout
the simulations presented here, growth effects due to a non-linear
focusing force may not appear.
The actually obtained emittance changes must therefore be
attributed to either a variation of the beam's ``free field energy''
as described by Eq.~(\ref{wangl}), or the action of Langevin forces
according to Eq.~(\ref{epsgro}).

In our simulations, equivalent ensembles of
macro-particles representing equivalent beams are tracked
through three distinct fictitious external focusing geometries.
At a time, the external focusing forces define (i)
a system that isotropically focuses the beam in all three
spatial directions, with the focusing forces acting
continuously along the beam line;
(ii) a system that again isotropically focuses the beam in all three
spatial directions, but with focusing forces now acting
periodically along the beam line; and (iii)
a system that focuses the beam anisotropically in the three
spatial directions, with focusing forces acting also
periodically along the beam line.

For the isotropic focusing systems, the simulations are launched
with both isotropic as well as anisotropic mismatch conditions.
The strength of a particular mismatch will be quantified by the
dimensionless ``mismatch factor'' $\Delta$~\cite{guyard}, defined as
\begin{equation}\label{mmf}
\Delta=\sum_{i=1}^{3}\left[{(\alpha_{i}^{\rm cs}-\alpha_{i,m}^{\rm cs})}^{2} -
(\beta_{i}^{\rm cs}-\beta_{i,m}^{\rm cs})
(\gamma_{i}^{\rm cs}-\gamma_{i,m}^{\rm cs})\right]\,,
\end{equation}
with $\alpha_{i}^{\rm cs}$, $\beta_{i}^{\rm cs}$, and
$\gamma_{i}^{\rm cs}$ denoting the Courant-Snyder~\cite{cousny}
functions for the actual and the matched beam, respectively.

\begin{figure}[ht]
\begin{center}
\epsfig{file=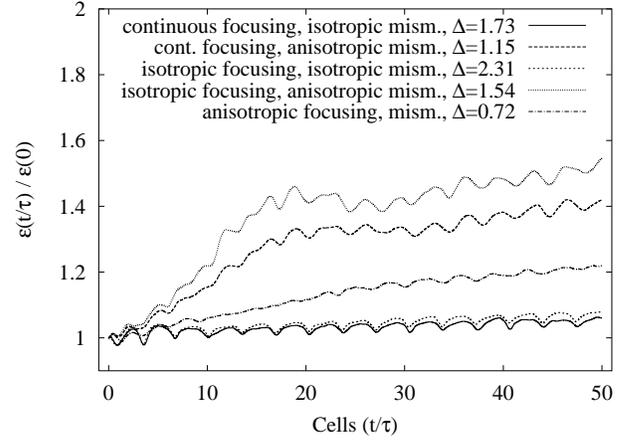,height=86mm,angle=-90}
\vspace{3mm}
\caption{Emittance growth factors $\varepsilon(t/\tau)/\varepsilon(0)$
for 5000 simulation particles versus number of cells obtained
by 3-D simulations of different focusing systems and matching
conditions at $\sigma_0=60^\circ$, $\sigma=15^\circ$ per cell.}
\label{fig:eps-5000}
\end{center}
\end{figure}

\begin{figure}[ht]
\begin{center}
\vspace{-5mm}
\epsfig{file=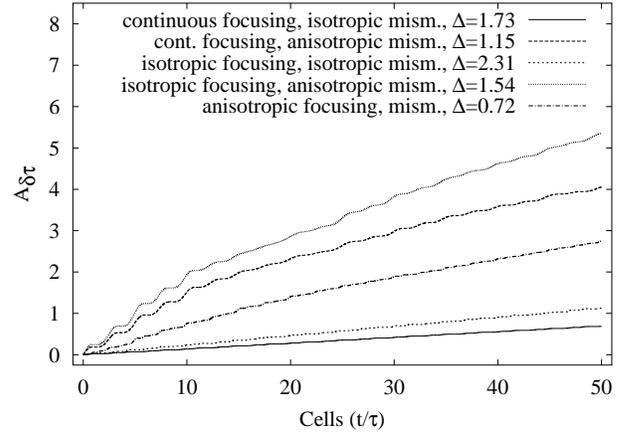,height=86mm,angle=-90}
\vspace{3mm}
\caption{Temperature anisotropy coefficients $A_{\delta\tau}$
for 5000 simulation particles versus number of cells obtained
by 3-D simulations of different focusing systems and matching
conditions at $\sigma_0=60^\circ$, $\sigma=15^\circ$ per cell,
for a normalized correlation time $\delta\tau/\tau=0.455$.}
\label{fig:aniso-5000}
\end{center}
\end{figure}

In Fig.~\ref{fig:eps-5000}, we plot the RMS emittance growth factors
obtained from the simultaneous numerical integration of $N$
coupled single particle equations of motion~(\ref{deteq})
with $N=5000$ macro-particles representing the beam.
We observe that the emittance growth rates even for strong
mismatch are much smaller if the beam stays isotropic --- as given
for isotropic external focusing in case of isotropic mismatch.
On the other hand, if the beam's phase space symmetry is rendered
anisotropic because of anisotropic focusing or mismatch, the
emittance growth factors come out considerably larger.
This behavior can be explained considering Fig.~\ref{fig:aniso-5000}.
It shows the evolution of the respective temperature anisotropy
coefficients $A_{\delta\tau}(t)$, defined by Eq.~(\ref{aniso}).
As is easily understood by their definition, the anisotropy
coefficients are much smaller for isotropic beam mismatch oscillations,
compared to anisotropic oscillations of comparable strength $\Delta$.
According to Eq.~(\ref{epsgro}), the anisotropy coefficients are directly
related to the irreversible part of the RMS emittance growth.
We therefore expect the actual emittance growth rates to follow the
magnitude of the temperature anisotropy if emittance changes due to
variations of the ``free field energy'' can be neglected.
This is indeed what we observe in our simulations.
As stated above, the friction factor $\gamma_{1}$ contained in
Eq.~(\ref{epsgro}) provides us with a global measure for the
magnitude of the Langevin forces that act within the
ensemble of beam particles.
We conclude that $\gamma_{1}$ should be similar for equivalent
ensembles of macro-particles, as defined in our simulations.

\begin{figure}[ht]
\begin{center}
\vspace{-5mm}
\epsfig{file=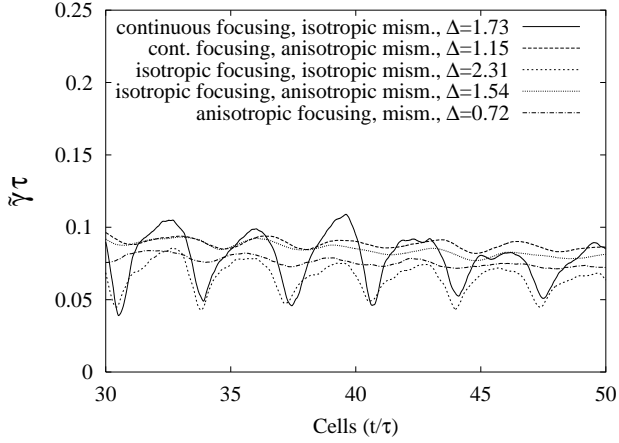,height=86mm,angle=-90}
\vspace{3mm}
\caption{Normalized friction coefficients $\tilde{\gamma}\tau$
for 5000 simulation particles versus number of cells obtained
by 3-D simulations of different focusing systems and matching
conditions at $\sigma_0=60^\circ$, $\sigma=15^\circ$ per cell,
for a normalized correlation time $\delta\tau/\tau=0.455$.}
\label{fig:gammatau-5000}
\end{center}
\end{figure}

We now estimate the normalized friction coefficient
$\gamma_{1}\tau$ that is consistent with the emittance growth
effects experienced in the simulations for the defined variety
of temperature anisotropies.
According to Eq.~(\ref{epsgro}), $\gamma_{1}$ determines the
amount of {\em irreversible\/} emittance growth that is obtained
for a given temperature anisotropy.
Since the {\em actual\/} emittance growth factors resulting from the
evolution of the simulated particle ensemble reflect a mixture of
reversible as well as irreversible effects, we cannot extract
$\gamma_{1}$ directly from the simulation data.
Instead, we plot in Fig.~\ref{fig:gammatau-5000} the related
factors $\tilde{\gamma}$, defined by
\begin{equation}\label{gammatau}
\tilde{\gamma} = \ln\frac{\varepsilon(t)}{\varepsilon(0)}\;\bigg/
\frac{t}{M}\sum_{m=1}^{M} A_{\delta\tau}(m\,\delta\tau)\,.
\end{equation}
Under the condition that emittance changes due to variations of the
``free field energy'' can be neglected, hence that only the
Langevin forces account for emittance changes, $\tilde{\gamma}$
is identical with the global friction factor $\gamma_{1}$.
Otherwise, only the time average of $\tilde{\gamma}$ provides us
with an approximation of $\gamma_{1}$.
With regard to Fig.~\ref{fig:gammatau-5000}, we convince ourselves
that this is indeed true --- at least to good approximation.
\vspace{-3mm}
The large fluctuations observed for the cases of isotropic beam symmetry
indicate that ``free field energy'' contributions to the emittance
change according to Eq.~(\ref{wangl}) take place.
Following from a linear perturbation analysis of the envelope
equations~\cite{bopa}, the phase advance $\sigma_{{\rm env},H}$ of
an isotropic beam ``breathing'' mode can be expressed in
terms of the zero current single particle phase advance $\sigma_{0}$
and the related phase advance $\sigma$ occurring in presence of space
charge forces as
\begin{equation}\label{mmosc}
\sigma_{{\rm env},H} = \sqrt{3\sigma_{0}^{2}+\sigma^{2}}\,.
\end{equation}
\begin{figure}
\begin{center}
\vspace{-5mm}
\epsfig{file=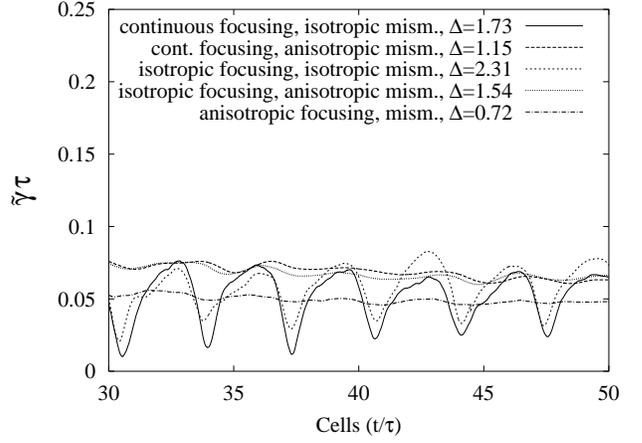,height=86mm,angle=-90}
\vspace{3mm}
\caption{Normalized friction coefficients $\tilde{\gamma}\tau$
for 10000 simulation particles versus number of cells obtained
by 3-D simulations of different focusing systems and matching
conditions at $\sigma_0=60^\circ$, $\sigma=15^\circ$ per cell,
for a normalized correlation time $\delta\tau/\tau=0.4$.}
\label{fig:gammatau-10000}
\end{center}
\end{figure}
As displayed in Fig.~\ref{fig:eps-5000} and Fig.~\ref{fig:gammatau-5000},
we observe 5 beam core oscillations in about 17 focusing periods,
which corresponds to a phase advance of about
$\sigma_{{\rm env},H}=106^{\circ}$ per cell for this mode.
This number is in excellent agreement with Eq.~(\ref{mmosc}),
predicting a value of $\sigma_{{\rm env},H}=105^{\circ}$ for our
simulation parameters.

The statement that the friction coefficient $\gamma_{1}$
is related to the magnitude of stochastic forces is confirmed
by simulations performed with different number of particles
while leaving all other simulation parameters invariant.
Fig.~\ref{fig:gammatau-10000} shows the normalized
friction coefficients $\tilde{\gamma}\tau$ obtained
from tracking of 10000 simulation particles.
A comparison with the corresponding Fig.~\ref{fig:gammatau-5000}
--- displaying the results for 5000 particles --- shows that the
$\tilde{\gamma}\tau$ values are reduced.
\begin{figure}[ht]
\begin{center}
\vspace{-5mm}
\epsfig{file=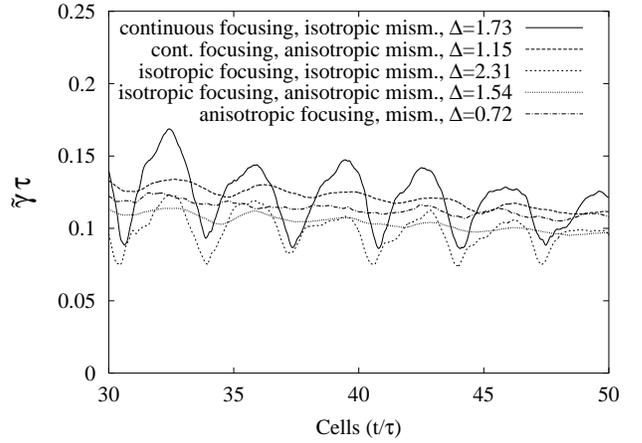,height=86mm,angle=-90}
\vspace{3mm}
\caption{Normalized friction coefficients $\tilde{\gamma}\tau$
for 2500 simulation particles versus number of cells obtained
by 3-D simulations of different focusing systems and matching
conditions at $\sigma_0=60^\circ$, $\sigma=15^\circ$ per cell,
for a normalized correlation time $\delta\tau/\tau=0.475$.}
\label{fig:gammatau-2500}
\end{center}
\end{figure}
Finally, Fig.~\ref{fig:gammatau-2500} shows the normalized friction
coefficients $\tilde{\gamma}\tau$ obtained for 2500 simulation particles.
Under these circumstances the $\tilde{\gamma}\tau$ values come out
larger compared to the previously cited cases, as expected.
As a rough estimate, we find that $\gamma_{1}$ scales with
the inverse square root of the number $N$ of simulation particles
\begin{displaymath}
\gamma_{1} \propto N^{-1/2}\,.
\end{displaymath}
\section{Conclusions}\label{sec:conclusions}
In this paper, we have reviewed the analytical description of
emittance growth effects that are caused by the actual
granularity of the charge distribution of particle beams ---
commonly referred to as intrabeam scattering.
Within the same framework, the description of noise phenomena that
necessarily accompany computer simulations of beams has been outlined.
A formula has been derived that relates the expected
emittance growth rate to both the magnitude of the noise force,
and the average temperature anisotropy the beam experiences within a
correlation time span $\delta\tau$.
We have presented computer simulations of beam transport through
various focusing lattices and matching conditions, each of them
enforcing a specific beam temperature anisotropy.
It has been shown that the numerically obtained emittance growth
rates can indeed be explained on the basis of this formula.
The magnitude of the noise force has been shown to depend
significantly on the number of simulation particles.
As a consequence, we may identify the related emittance growth
rates as ``computer noise'' artifacts.

Discreteness errors inevitably emerge in computer simulations
of dynamical systems.
Therefore, the actual time evolution of the simulated
system always encloses irreversible aspects --- even if the
actually coded equations of motion are strictly reversible.
In that sense, the simulation results can be regarded as exact solutions
of a modified dynamical system that always comprises Langevin force terms.
The magnitude of these ``computer noise'' related forces strongly depends
on the specific realization of the simulation.
The simulation may thus pretend effects that either do not
occur at all within the ``real'' system in question,
or that occur at different time scales.

The crucial point for the correct interpretation of computer simulations
of beam dynamical systems is to keep in mind that the appearance of
noise-related emittance growth depends on both the magnitude of the
noise forces as well as the time averaged temperature anisotropy.
Therefore, macroscopic emittance growth effects do not appear in cases
where the temperature anisotropy within the system is negligible ---
even if the ``computer noise'' related forces are large.
On the other hand, even if strictly periodic solutions of
non-autonomous Vlasov-Poisson systems exist, the results of computer
simulations of such systems will never be strictly periodic.
Only if we take into account these subtle differences, we avoid
misinterpretations of our computer simulation results.
\acknowledgments%
The author is indebted to C.~Riedel and I.~Hofmann from GSI for the
many valuable discussions during the writing of this paper.
%\end{multicols}


\begin{thebibliography}{99}
%\begin{multicols}{2}
\bibitem[]{kapvla}
  I.M.~Kapchinskij and V.V.~Vladimirskij,
  {\it Proceedings of the International Conference on High Energy
  Accelerators, CERN, Geneva, 1959}, (CERN, Geneva, 1959), p.~247.
\bibitem[]{strklarei}
  J.~Struckmeier, J.~Klabunde, and M.~Reiser,
  Part.~Accel.~{\bf 15}, 47 (1984).
\bibitem[]{wangler}
  T.P.~Wangler, K.R.~Crandall, R.S.~Mills, and M.~Reiser,
  IEEE~Trans.~Nucl.~Sci.~{\bf 32}, 2196 (1985).
\bibitem[]{reiser}
  M.~Reiser, {\it Theory and Design
    of Charged Particle Beams\/} (Wiley, New York, 1994).
\bibitem[]{hola}
  I.~Hofmann, L.J.~Laslett, L.~Smith, and I.~Haber,
  Part.~Accel.~{\bf 13}, 145 (1983).
\bibitem[]{hotemp}
  I.~Hofmann,
  {\it Advances in Electronics and Electron Physics}, Supplement 13C,
  edited by A.~Septier, (Academic Press, New York, 1983), pp.~49--140.
\bibitem[]{birdsall}
  C.K.~Birdsall, G.R.~Brewer, and A.V.~Haeff,
  Proc.~IRE~{\bf 41}, 865 (1953).
\bibitem[]{neil}
  V.K.~Neil and A.M.~Sessler,
  Rev.~Sci.~Instrum.~{\bf 36}, 429 (1965).
\bibitem[]{strrei}
  J.~Struckmeier, and M.~Reiser,
  Part.~Accel.~{\bf 14}, 227 (1984).
\bibitem[]{hobson}
  A.~Hobson, {\it Concepts in Statistical Mechanics\/}
  (Gordon and Breach Science Publishers, New York, 1971).
\bibitem[]{piwi}
  A.~Piwinski, {\it Proceedings of the  9th International Conference
  on High Energy Accelerators, Stanford, 1974}, (SLAC, Stanford, 1974), p.~405.
\bibitem[]{chandra}
  S.~Chandrasekhar,
  Rev.~Mod.~Phys.~{\bf 15}, 1--89 (1943).
\bibitem[]{biso}
  J.~J.~Bisognano, in {\it Physics of High Energy Particles, Stony Brook, 1983},
  edited by M.~Month, P.~F.~Dahl, and M.~Dienes, AIP~Conf.~Proc.~No.~127
 (AIP, New York, 1985), p.~443.
\bibitem[]{struck-pa}
  J.~Struckmeier,
  Part.~Accel.~{\bf 45}, 229 (1994).
\bibitem[]{struck}
  J.~Struckmeier,
  Phys.~Rev.~E~{\bf 54}, 830 (1996).
\bibitem[]{jowett}
  J.~M.~Jowett, in {\it Physics of Particle Accelerators, SLAC, Stanford, 1985},
  edited by M.~Month and M.~Dienes, AIP~Conf.~Proc.~No.~153
  (AIP, New York, 1987), p.~864.
\bibitem[]{langevin}
  P.~Langevin,
  Comptes rendus~{\bf 146}, 530 (1908).
\bibitem[]{haber}
  I.~Haber, D.~A.~Callahan, C.~M.~Celata, W.~M.~Fawley, A.~Friedman,
  D.~P.~Grote, and A.~B.~Langdon, in {\it Space Charge Dominated Beams
  and Applications of High Brightness Beams, Bloomington, 1995},
  edited by S.~Y.~Lee, AIP~Conf.~Proc.~No.~377,
  (AIP, New York, 1995), p.~244.
\bibitem[]{flekk}
  E.G.~Flekk\o{}y and P.V.~Coveney,
  Phys.~Rev.~Lett.~{\bf 83}, 1775 (1999).
\bibitem[]{kramers}
  H.A.~Kramers,
  Physica~{\bf 7}, 284 (1940).
\bibitem[]{moyal}
  J.E.~Moyal,
  J.~R.~Stat.~Soc.~(London)~B~{\bf 11}, 150 (1949).
\bibitem[]{risken}
  H.~Risken, {\it The Fokker-Planck Equation\/}
  (Springer, Berlin, Heidelberg, New York, 1989).
\bibitem[]{lund}
  S.M.~Lund, J.J.~Barnard, and J.M.~Miller,
  {\it Proceedings of the 1995 Particle Accelerator Conference, Dallas, 1995},
  (IEEE, Piscataway, NJ, 1996) p.~3278.
\bibitem[]{liboff}
  R.L.~Liboff,
  {\it Kinetic Theory\/} (John Wiley \& Sons, New York, 1998).
\bibitem[]{jansen}
  G.H.~Jansen,
  {\it Coulomb Interaction in Particle Beams\/}
  (Academic Press, New York, 1990).
\bibitem[]{shannon}
  C.E.~Shannon,
  {\it The Mathematical Theory of Communication\/}
  (University of Illinois Press, Urbana, IL, 1949).
\bibitem[]{host}
  I.~Hofmann and J.~Struckmeier,
  Part.~Accel.~{\bf 21}, 69 (1987).
\bibitem[]{pacc1}
  J.~Struckmeier and I.~Hofmann,
  Part.~Accel.~{\bf 39}, 219 (1992).
\bibitem[]{guyard}
  J.~Guyard and M.~Weiss,
  {\it Proceedings of the 1976 Proton Linear Accelerator Conference,
  Chalk River}, edited by S.~O.~Schriber
  (Atomic Energy of Canada Ltd., Chalk River, 1976), p.~254.
\bibitem[]{cousny}
  E.D.~Courant and H.S.~Snyder,
  Ann.~Phys.~(N.Y.) {\bf 3}, 1--48 (1958).
\bibitem[]{bopa}
  A.~Letchford, K.~Bongardt, and M.~Pabst,
  {\it Proceedings of the 1999 Particle Accelerator Conference, New York},
   (IEEE, Piscataway, NJ, 1999), p.~1767.
%\end{multicols}
\end{thebibliography}
\end{document}